\documentclass[twocolumn,english,aps,pre,floats]{revtex4}
\usepackage[T1]{fontenc}
\usepackage[latin9]{inputenc}
\usepackage[pdftex]{graphicx}
\usepackage{amsmath}
\usepackage{amssymb}

\makeatletter
\@ifundefined{textcolor}{}
{%
 \definecolor{BLACK}{gray}{0}
 \definecolor{WHITE}{gray}{1}
 \definecolor{RED}{rgb}{1,0,0}
 \definecolor{GREEN}{rgb}{0,1,0}
 \definecolor{BLUE}{rgb}{0,0,1}
 \definecolor{CYAN}{cmyk}{1,0,0,0}
 \definecolor{MAGENTA}{cmyk}{0,1,0,0}
 \definecolor{YELLOW}{cmyk}{0,0,1,0}
 }

\usepackage{latexsym}
\usepackage{amsbsy}
\@ifundefined{definecolor}
 {\@ifundefined{definecolor}
 {\usepackage{color}}{}
}{}

\@ifundefined{definecolor}
 {\@ifundefined{definecolor}
 {\usepackage{color}}{}
}{}

\usepackage{multirow}

\makeatother

\usepackage{babel}

\makeatother

\usepackage{babel}

\begin{document}

\title{Network-based confidence scoring system for genome-scale metabolic
reconstructions}

\author{M. \'Angeles Serrano and Francesc Sagu\'es}

\affiliation{Departament de Qu\'{\i}mica F\'{\i}sica, Universitat de Barcelona, Mart\'{\i} i
Franqu\`es 1, 08028 Barcelona, Spain}
\begin{abstract}
Reliability on complex biological networks reconstructions remains
a concern. Although observations are getting more and more precise,
the data collection process is yet error prone and the proofs display
uneven certitude. In the case of metabolic networks, the currently
employed confidence scoring system rates reactions according to a
discretized small set of labels denoting different levels of experimental
evidence or model-based likelihood. Here, we propose a computational
network-based system of reaction scoring that exploits the complex
hierarchical structure and the statistical regularities of the metabolic
network as a bipartite graph. We use the example of \textit{Escherichia
coli} metabolism to illustrate our methodology. Our model is adjusted
to the observations in order to derive connection probabilities between
individual metabolite-reaction pairs and, after validation, we integrate
individual link information to assess the reliability of each reaction
in probabilistic terms. This network-based scoring system breaks the
degeneracy of currently employed scores, enables further confirmation
of modeling results, uncovers very specific reactions that could be
functionally or evolutionary important, and identifies prominent experimental
targets for further verification. We foresee a wide range of potential
applications of our approach given the natural network bipartivity
of many biological interactions. 
\end{abstract}
\maketitle


\section{Introduction}

Cells are self-organized entities able to carry out specialized functions
at several interrelated levels, i.e. a genetic code is repeatedly
transcribed for cell maintenance and reproduction, a proteomic set
robustly embodies the cell machinery, and a mesh of metabolic reactions
continuously furnishes the energy and biochemical compounds necessary
for life. A crucial milestone to understand and control cellular behavior
is the building up of reliable reconstructions of the interactions
spanning the different functional levels~\cite{Kuhner:2009,Yus:2009,Guell:2009}.
Such reconstructions find a natural abstraction in the form of complex networks~\cite{Newman:2003,Dorogovtsev:2008,Barrat:2008}, where nodes represent cellular components, such as genes,
proteins or metabolites, while edges identify the presence of biological
interactions between them~\cite{Aittokallio:2006}. These network
representations enable to map the large-scale structure of cellular
interactions~\cite{Barabasi:2004,Kepes:2007}, to explore the basic
principles of transcriptome and proteome organization~\cite{Kepes:2007}, to identify missing genes encoding for specific metabolic functions~\cite{Kharchenko:2006},
and to analyze emergent global phenomena in metabolism like robustness
and regulation~\cite{Stelling:2002,Guimera:2005b,Motter:2008,Smart:2008}.

At present, the information for complex network representations of
cellular systems comes primarily from web-based databases, oftentimes
manually curated with information from multiple sources, like annotations
from the literature or new experiments~\cite{Palsson:2006}. It is
common to take the reliability of these data for granted and to draw
from them resolute inferences about the properties or the behavior
of the investigated organisms~\cite{Oberhardt:2009}. Although, in general, observations
are getting more and more precise, uncertainties about components
or interactions remain~\cite{Amaral:2008,Kaltenbacha:2009}: experimental
targets are many times biased towards the most rewarding in terms
of expected impact, experimental evidence gathered with different
methodologies is not always of the same quality, variability is unavoidably
present in different organisms of the same species, and perfect environmental
control in experiments is often difficult to achieve. In particular,
high-throughput techniques produce massive data in comparison with
more dedicated experiments at the price of repeatedly reported inaccuracy~\cite{Mering:2002}.
In the best case, a number of alternative or missing interactions
need to be inferred and those detected with low confidence need to
be validated.

Prediction of missing interactions in probabilistic terms is possible
on the basis of the structure of the network alone, and could serve
to better characterize network-based descriptions of biological systems
and as a guide for new experiments. Despite being of enormous importance,
this task is just a first step towards assessing the quality of a
given data set. Metabolisms, in particular, encode information through
the metabolic processes themselves, either singled out or combined
in pathways. It would be thus convenient to assess the realiability
of a reconstructed metabolic network in terms of a reaction-based
{}``confidence scoring system''. The latter, although it may appear
at first sight {}``chemically spurious'' (canonically a chemical
reaction either exists or does not), is fully admitted in the context
of prone-to-error large-scale reconstructed biochemical networks~\cite{Francke:2005,Thiele:2010}.
In short, it amounts to assign a certain level of experimental evidence,
or model-based likelihood, to every particular reaction within a genome-scale
metabolic reconstruction.

For network analysts, scoring a biochemical process in terms of its
confidence, apart from being experimentally motivating, is a computational
challenge. Starting from the assessment of individual links, as it
has been devised for instance for protein-protein interaction networks~\cite{Saito:2002,Goldberg:2003},
the problem is conditioned by the need of non-local models that faithfully
capture the large-scale statistical regularities of the networks,
and by the requirement of sufficient reliability in the input experimental
observations. In this respect, the reconstruction of genome-scale
biochemical networks is a well-established procedure down to stoichiometric
detail and available metabolic network reconstructions, such as those
of \textit{E. coli}, are of sufficient quality and can be exploited
to get accurate network-based predictions.

Here, we propose a computational network-based system of reaction
scoring and, subsequently, apply it to the metabolism of \textit{E.
coli}. To this end, we introduce a link prediction method that exploits
the complex hierarchical structure and the statistical regularities
of the metabolic network~\cite{Wagner:2001,Ravasz:2002} and takes
explicitly into account its bipartite nature~\cite{Newman:2001b,Holme:2003}.
Our model is adjusted to the observations in order to derive connection
probabilities between pairs of elements~\cite{Serrano:2008a,Clauset:2008,Guimera:2009}
and, after validation, we integrate individual metabolite-reaction
link information to assess the reliability of each reaction in the
metabolism of \textit{E. coli}, both in absolute terms and relative
to a random null model. Such, for the first time proposed, network-based
scoring system permits to rank reactions in the database, and compare
our {}``in silico'' predictions with {}``in vivo'' annotations
quoted in the reference literature \cite{Thiele:2010}. In this latter
context, confidence scores are discretized according to five integers.
From a direct comparison with our continuously distributed index,
similarities and discrepancies are fleshed out, singularly stressing
the specifities of particular reactions as compared to those that
are baseline. At the end, our versatile approach may provide unexisting
information for unranked databases, or refine and complement labeled
ones, aiming to direct more effectively experimental efforts and to
unveil structural levels of organization within the intrincate biological
networks underlying living organisms.
\begin{figure}[t]
\hspace{-0.7cm}
\includegraphics[width=8.7cm]{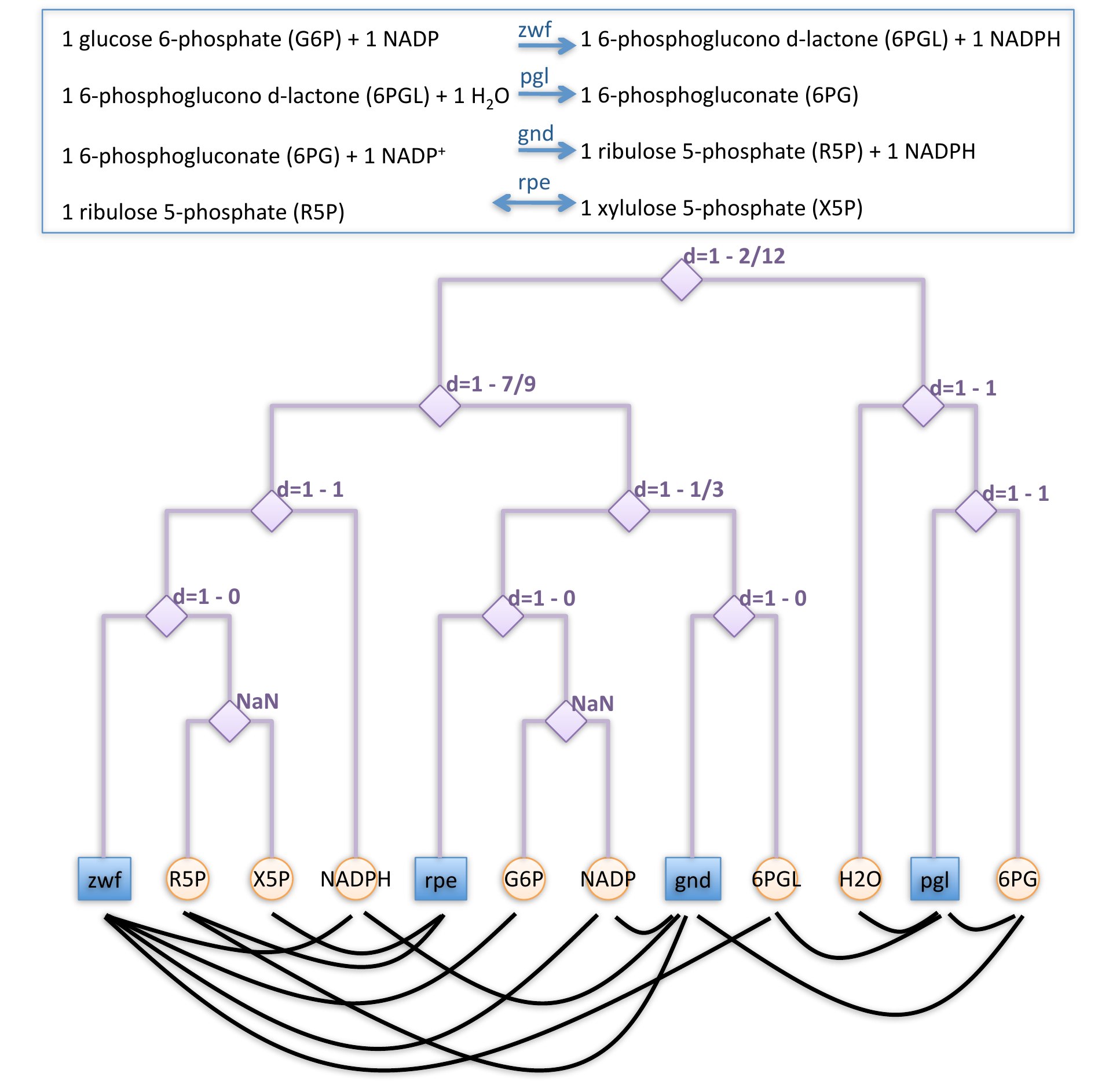} 
\caption{\textbf{Bipartite network representation and corresponding hierarchical
dendrogram.} We illustrate the hierarchical random graph model with
four coupled stoichiometric equations in the pentose-phosphate pathway
of \textit{E. coli}, taken from~\cite{Wagner:2001}, represented as a bipartite network. Reaction acronyms
stand for the catalyzing enzyme: zwf, \textit{glucose- 6- phosphate dehydrogenase}
{[}EC 1. 1.1. 49{]} ; pgl, \textit{6- phospho-gluconolactonase} {[}EC 3. 1.
1.31{]} ; gnd, \textit{6- phosphogluconate dehydrogenase} {[}EC 1.1.1. 43{]}
; rpe, \textit{ribulose- phosphate 3- epimerase} {[}EC 5. 1.3. 1{]}. These
equations are represented as an unweighted undirected bipartite network
formed by connections (black lines) between reactions (blue squares)
and metabolites (orange circles) existing whenever one metabolite
participates in a reaction. Notice that metabolites or reactions are
not connected among themselves. The model adjusts an underlying binary
tree to the observed network structure where each internal node $t$
is associated with a tree probability $\rho_{t{\normalcolor }}$ that
we transform into a distance $d_{t}=1-\rho_{t}$. Every pair reaction-metabolite
with minimum common ancestor $t$ is then assumed to be separated
a distance $d_{mr}=d_{t}$ (the $NaN$ notation indicates that only
one class of nodes populates the leaves of the child branches and
are thus not considered).}
\label{fig:hrgb4} 
\end{figure}

\section{Results and Discussion}
\subsection{Network-based confidence scoring system}
Beyond experimental evidence, it is possible to assess confidence
scores for reactions in genome-scale metabolic reconstructions using
theoretical models. Given an experimentally derived metabolic reconstruction,
a confidence score for the reactions can be computed on the basis
of a suited stochastic network-based model, as proposed in next section.
In the context of bipartite network representations~\cite{Newman:2001b,Holme:2003},
the model exploits the statistical regularities that imprint the structure
of the metabolic network in order to ascertaining how well individual
links between metabolites and reactions fit the observed topological
patterns. In this way, it is possible to predict probabilities of
connection for all potential interactions, those already present in
the reconstruction and those absent. These probabilities are further
integrated for the specific combination of metabolites involved in
any particular reaction.

To define this network-based confidence score in quantitative terms,
we interpret that a reaction is equivalent to the univocal combination
of its associated metabolites, and that every metabolite $m$ has
a probability $p_{mr}$ of being associated to a reaction $r$. Once
the connection probabilities between all metabolites and all reactions
are estimated from the model, and assuming their mutual independence,
the probability that a certain combination $\mu$ of metabolites co-occur
in a reaction $r$ is

\begin{equation}
p_{\mu r}=\prod_{m\in\mu}p_{mr}\prod_{m'\notin\mu}(1-p_{m'r}),\label{nur}\end{equation}
 where the subindex $m$ corresponds to metabolites in the predefined
set $\mu$ and $m'$ to those not included. Notice that in general
the set $\mu$ could be different from the actual combination of metabolites
associated to reaction $r$. The average number of co-occurrences can be calculated as the sum
over all reactions \begin{equation}
n_{\mu}=\sum_{r=1}^{R}p_{\mu r}.\label{ssc}\end{equation}
 A network-based confidence score for a reaction can then be defined
as the average number of occurrences $n_{\nu}$ that the model associates
to its particular combination of metabolites $\nu$. Defined in this
way, these scores break the degeneracy of reactions with identical
experimental evidence.

\subsection{The Tree Distance Bipartite (TDB) model}
In the following, we introduce and discuss a stochastic network-based
model to estimate the probabilities $p_{mr}$ of connection between
metabolites and reactions. Taking advantage of their natural bipartite
nature, we consider network representations where both metabolites
and reactions appear as nodes and metabolites are connected by edges
to the reactions they participate in. We consider the simplest unweighted
undirected representation, where substrates and products are not differentiated.
See Fig.~1 for an example.

Previous works have shown that the complex organization of metabolic networks
displays characteristic features shared by other complex networks:
short topological diameter~\cite{Ma:2003a}, steady state cycles~\cite{Wright:2008}
or structural robustness~\cite{Smart:2008}, for instance. We implement
a large-scale model that takes advantage of some of those organizing
principles, in particular the heterogeneity in the number of connections
per metabolite (degree) ~\cite{Jeong:2000} and its hierarchical
architecture~\cite{Ravasz:2002}, to infer connection probabilities
between metabolites and reactions. Network-based models are usually
prescribed in terms of connection rules between the nodes. These laws
are stated independently of observed systems to produce graphical
representations that summarize their topological structure. Notice
that here, in contrast, we compute from the observed data the set
of connection probabilities that has the maximum likelihood to reproduce
the structure of an experimental metabolic reconstruction, so we are
solving the inverse problem.

Our first step is to assume an underlying metric space for the metabolic
network where all its elements are supposed that the closer they are,
the higher will be their connection probability. To this end, we fit
a hierarchical random graph~\cite{Clauset:2008} to the metabolic
reconstruction, once represented as a bipartite network with $M$
metabolites and $R$ reactions (see Fig.~1). More specifically, we
adjust the observed bipartite network to a dendrogram, or binary tree
structure $T$, where metabolites and reactions appear as distinguishable
leafs. This tree represents the underlying metric space, and each
of the $M+R-1$ internal nodes $t$ in the dendrogram has an associated
distance $d_{t}$, so that each pair metabolite-reaction for which
$t$ is the lowest common ancestor is separated by a distance in the
tree $d_{mr}=d_{t}$, independently from whether the link actually
exists in the network or not. We find these tree distances by fitting
the tree to the observed network data combining a maximum-likelihood
approach with a Monte Carlo sampling method that explores the space
of all possible dendrograms (see Appendix A). Our results
are based on intensive numerical simulations that average a large
number of samples in the stationary state when changes in the form
of the dendrogram do not modify the likelihood function beyond fluctuations.

Once a distance $d_{mr}$ is associated to every possible pair metabolite-reaction,
we correct for heterogeneity in the degrees of metabolites and compute
the connection probability between metabolite $m$ and reaction $r$
as \begin{equation}
p_{mr}=\frac{1}{1+\frac{d_{mr}R}{k_{m}}},\label{prenor}\end{equation}
 where $k_{m}$ is the degree of the metabolite. As a result, our
model produces a list of estimated connection probabilities $p_{mr}$
between all possible combinations metabolite-reaction. The confidence
score for every specific reaction can then be computed by applying
Eq.~(\ref{nur}) and Eq.~(\ref{ssc}).
\begin{figure}[t]
\hspace{-0.7cm}
\includegraphics[width=8.7cm]{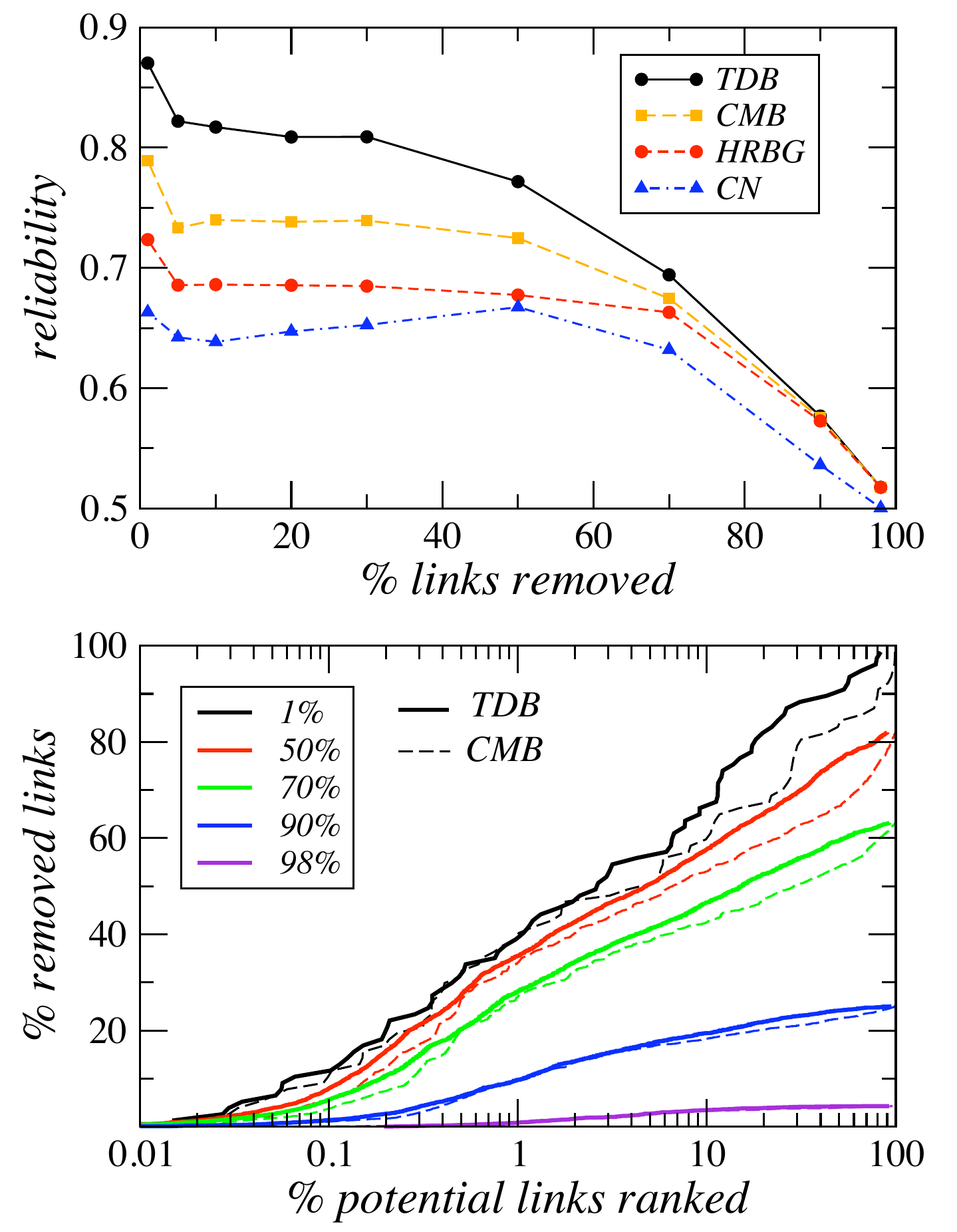} 
\caption{\textbf{Evaluation of the reliability of the Tree Distance Bipartite
model. Top}. Link reliability is calculated in statistical terms as
the probability that a randomly chosen removed connection has higher
estimated connection probability by our method than a randomly chosen
pair of unconnected reaction-metabolite. The TDB model presented here
is compared against the CMB model, the HRBG model and the CN model
(see Appendix B). \textbf{Bottom}. Distribution of removed
links in the set of potential links ranked from more to less probable
according to the model. Notice that the higher the number of removed
links in the original network, the lower the total number of removed
links in the list of potential pairs. This is due to the fact that
the removal of all the links of a node is equivalent to the removal
of the node itself and then it does not add to the list of potential
connections.}
\label{fig:accuracy} 
\end{figure}

\subsection{TDB confidence scores for {\textit E. coli} metabolism}
As an application of this methodology, we analyze the iAF1260 version
of the K12 MG1655 strain of the metabolism of \textit{E. coli}~\cite{Feist:2007}
provided in the BIGG database (http://bigg.ucsd.edu/). From the empirical
data, we built a bipartite network representation (see Appendix C) avoiding reactions that do
not involve direct chemical transformation (we obviate isomerization,
diffusion and exchange reactions). This leads to a bipartite graph
of $1479$ reactions and $976$ metabolites. As expected, the number
of metabolites entering into a reaction $k_{r}$ follows a nearly
homogeneous distribution with mean $<k_{r}>=4.82$ and mode $5$.
In contrast, the number $k_{m}$ of reactions in which a metabolite
participates displays a scale free degree distribution $P(k_{m})\sim k_{m}^{-2.1}$,
with an average degree $<k_{m}>=7.30$ (see Appendix C).
Currency metabolites are the most connected substrates, some with
more than a hundred and up to $841$ connections ($h^{+}$, $h2o$,
$atp$, $pi$, $adp$, $ppi$, $nad$, $nadh$).

We obtain the TDB based probabilities $p_{mr}$ for this \textit{E.
coli} reconstruction, and we evaluate their reliability in statistical terms. We compute the
probability that the model associates a higher likelihood to a missing
link that has been removed from the bipartite metabolic network than
to a non-existing link that was never there. To this end, a subset
of links in the original network is removed uniformly at random and
a new set of connection probabilities is calculated on the basis of
the remaining part of the network. The new probabilities associated
to removed connections are compared one by one to that of non-existing
links. This reliability statistic ranges from $0.5$ to $1$ and indicates
how much better our method performs as compared to a by chance baseline
accuracy of $0.5$.

Figure 2 (top) shows the reliability index for different fractions
of links in the removed set. When $1\%$ of the $7127$ links in the
network are removed, it takes a value of $0.87$, meaning that $87\%$
of the times removed links are ranked higher in probability by our
model than non-existing links. In the same plot, we compared with
alternative models (see Appendix B), such as the Configuration
Model for Bipartite networks~\cite{Newman:2001b,Newman:2002c,Guillaume:2006}
(CMB), the Hierarchical Random Graph model~\cite{Clauset:2008} generalized
to bipartite networks (HRBG), and a local approach based on the computation
of Common Neighbors (CN)~\cite{Lorrain:1971} (reactions) between
pairs of metabolites. We find that our TDB model outperforms all other
strategies at identifying missing interactions. For each point in
Fig. 2 (top), we also measured the positions of the removed links
in the ranking of potential links ordered from more to less probable
according to the model. The results are shown in Fig. 2 (bottom).
It can be observed that the top $1\%$ of the predictions is highly
accurate in the sense that it contains nearly $40\%$ of all removed
connections. As expected, this value decreases as less links are available
to the algorithm in the remaining part of the network, but it is noticeable
that, even with a $70\%$ of the links removed from the original network,
the accumulation of those at the top $1\%$ as ranked by the algorithm
is still of more than $20\%$.
\begin{figure}[t]
\hspace{-0.7cm}
\includegraphics[width=8.7cm]{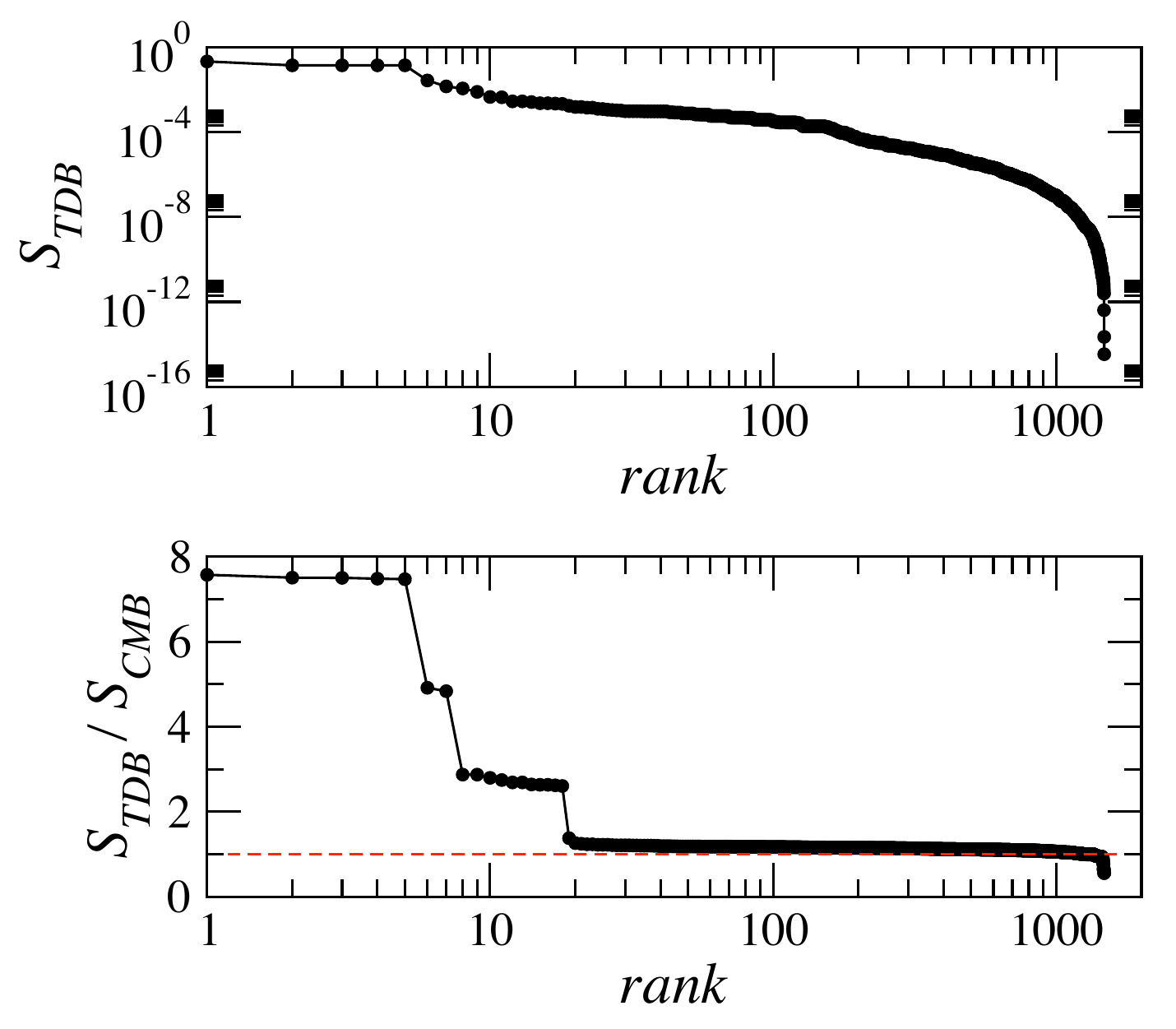} 
\caption{\textbf{TDB confidence scores and comparison with CMB confidence scores.
Top}. The TDB confidence scores $S_{TDB}$ have been ranked in decreasing
order. The higher values form a short plateau and correspond to outlier
values associated with reactions that only involve carrier metabolites
(hubs). \textbf{Bottom}. Ratio of TDB confidence scores $S_{TDB}$
to CMB confidence scores $S_{CMB}$, ranked in decreasing order and
compared to the baseline value equal to 1 (red dashed line). If $\Sigma=S_{TDB}/S_{CMB}>1\mbox{ }(<1)$
the network displays distance correlations (anticorrelations) which
are absent in the random case.}
\label{fig:scores} 
\end{figure}

In view of these results, we accept the accuracy at the statistical
level of the predicted probabilities $p_{mr}$ and we use them to
compute theoretical confidence scores $S_{TDB}=n_\nu(p_{mr}^{TDB})$ following Eq.~(\ref{nur})
and Eq.~(\ref{ssc}) (the detailed list of TDB confidence scores is available under request). Very high values of the TDB confidence
scores are typically associated to non-specific reactions dominated
by carrier metabolites (hubs in network-based terms). At the top of
the rank, the five reactions with the highest values form a group
of outliers (short first plateau in the top graph of Fig.~\ref{fig:scores})
corresponding to reactions whose metabolites are exclusively hubs.
These non-specific reactions are at the end of catabolic chains and
are likely to be shared by many different organisms. At the other
extreme of the spectrum, low values of $S_{TDB}$ are associated with
very specific reactions involving rare metabolites, in the sense that
they enter a small number of reactions. In between, $S_{TDB}$ scores
adopt a broad distribution of continuous values ranging several orders
of magnitude.

\subsection{Comparison beween Database (DB) and Tree Distance Bipartite (TDB)
scoring systems}
In the database, every reaction (except for exchanges and transports
through outer membrane) is endowed with a confidence score $S_{DB}$
assessing its level of evidence. These values are discrete and range
from 4 at the top, when there is direct biochemical proof, to 0 at
the bottom, when the reaction is included with no experimental evidence,
only because it improves modeling results. In between, values
of 3 correspond to genomic evidence, level 2 refer to sequence homology
evidence, and 1 stands for physiological evidence. This confidence
scoring system presents some shortcomings, one being the wicked level
of degeneracy implicit in the use of only five discrete values for
lists of hundreds or thousands of reactions.

Our TDB confidence scoring system breaks this degeneracy with a continuous
spectrum of values. We prefer to restrict to the extrema of our $S_{TDB}$
spectrum to compare against $S_{DB}$ scores. For reactions with scores
in the database that indicate a strong experimental evidence (values
$4$ or $3$), we find both high and low values of $S_{TDB}$ scores.
The first situation corresponds to our TDB model providing complementary
computational verification of experimental certainty. The second and
more interesting cathegory uncovers very specific reactions that could
be functionally or evolutionary important. Examples are the five \textit{FMNH2-dependent
monooxygenase} reactions, the \textit{Pyridoxine 5'-phosphate synthase}
reaction, or the \textit{Taurine dioxygenase} reaction. Conversely,
a weak experimental evidence, $S_{DB}$ scores $2$ or $1$, but a
high value of the $S_{TDB}$ score, qualifies the reaction as a good
target for further experimental verification in standard conditions. Many examples are found
within the transport subsystems, like reactions of transport via ABC
system (\emph{iron (II)} and \emph{(III)}, \emph{phosphatidylglycerol,
phosphatidate}). If the $S_{TDB}$ score is low, the reaction could be difficult to be observed experimentally except for very specific environments. Finally, high $S_{TDB}$ scores for reactions that
where required for modeling, $S_{DB}$ score $0$, denotes consistency
between our model and steady-state flux optimization solutions. It
is worth remarking that these reactions appear at the very end of
the catabolic chain and involve more than one carrier metabolites.
However, a variety of reactions, like many in the subsystem of \textit{Cofactor
and Prosthetic Group Biosynthesis} and many with the highest reaction
degree, manifest discrepancy between the models.

\subsection{Comparison beween Tree Distance Bipartite (TDB) and Configuration
Model Bipartite (CMB) confidence scores}
Along with absolute confidence scores, we also analyze relative scores
defined on the basis of the Configuration Model for Bipartite networks
(CMB)~\cite{Newman:2001b,Newman:2002c,Guillaume:2006} (see Appendix B). The latter assumes the actual degree distributions for
metabolites and reactions and it is otherwise maximally random in
the assembly of connections. For every reaction, we calculate the
$S_{CMB}$ score representing its probability of occurrence according
to the configuration model and use this value to compute the ratio
$\Sigma=S_{TDB}/S_{CMB}$ (the detailed list of relative scores is available under request). Since differences between both scores are mainly related to
the consideration of tree distances between metabolites and reactions
in the TDB model, a relative score $\Sigma=S_{TDB}/S_{CMB}>1\mbox{ }(<1)$
points to the presence of tree distance correlations (anticorrelations)
in the bipartite network, which are absent in the random case. In
other words, a ratio higher (smaller) than one for a certain reaction
indicates that its metabolites have a tendency to aggregate (avoid
each other) as compared to the random case. 

The ranking of relative scores is shown in the bottom graph of Fig.~\ref{fig:scores},
where several clusters can be differentiated. The first three clusters
appear in slightly tilted plateaus with levels well separated by appreciable
jumps. Each of them is formed by a subgroup of reactions that, according
to the database, tend to belong to the same subsystem and share characteristic
combinations of metabolites. The first group includes the \textit{FMNH2-dependent
monooxygenase} reactions, mentioned above as highly specific, with
\textit{flavin mononucleotide} and \textit{sulfite} as reactants. The
two reactions in the\textit{ }\textit{\emph{second plateau belong
to the }}\textit{Glycerophospholipid Metabolism} subsystem and are
the only two in the database associated to \textit{acyl phosphatidylglycerol}.
The third cluster gathers together the eleven reactions containing
\textit{2-Demethylmenaquinone 8}. It is remarkable that, in general,
the reactions in these clusters have attached a high DB confidence
score. Exceptions appear in the third cluster, where one reaction
has an experimental evidence score less than 3 while four reactions
are included for modeling reasons and protrude as experimental targets
for experimental verification. For the rest, most of the scores have
values above but close to one and there are also over two hundred
reactions with ratios below one. At the very tail, one finds a set
of reactions that share the common characteristic of being those with
the highest reaction degree and with weak or just modeling evidence.
In particular, \textit{thiazole phosphate synthesis} is the solely
reaction involving twelve metabolites in the database and has the
lowest relative score $\Sigma=0.3$, and noticeably also the lowest
absolute $S_{TDB}$ score ($S_{DB}=2$).

\subsection{Metabolic confidence map at the level of pathways}
\begin{figure}[t]
\hspace{-0.7cm}
\includegraphics[width=8.7cm]{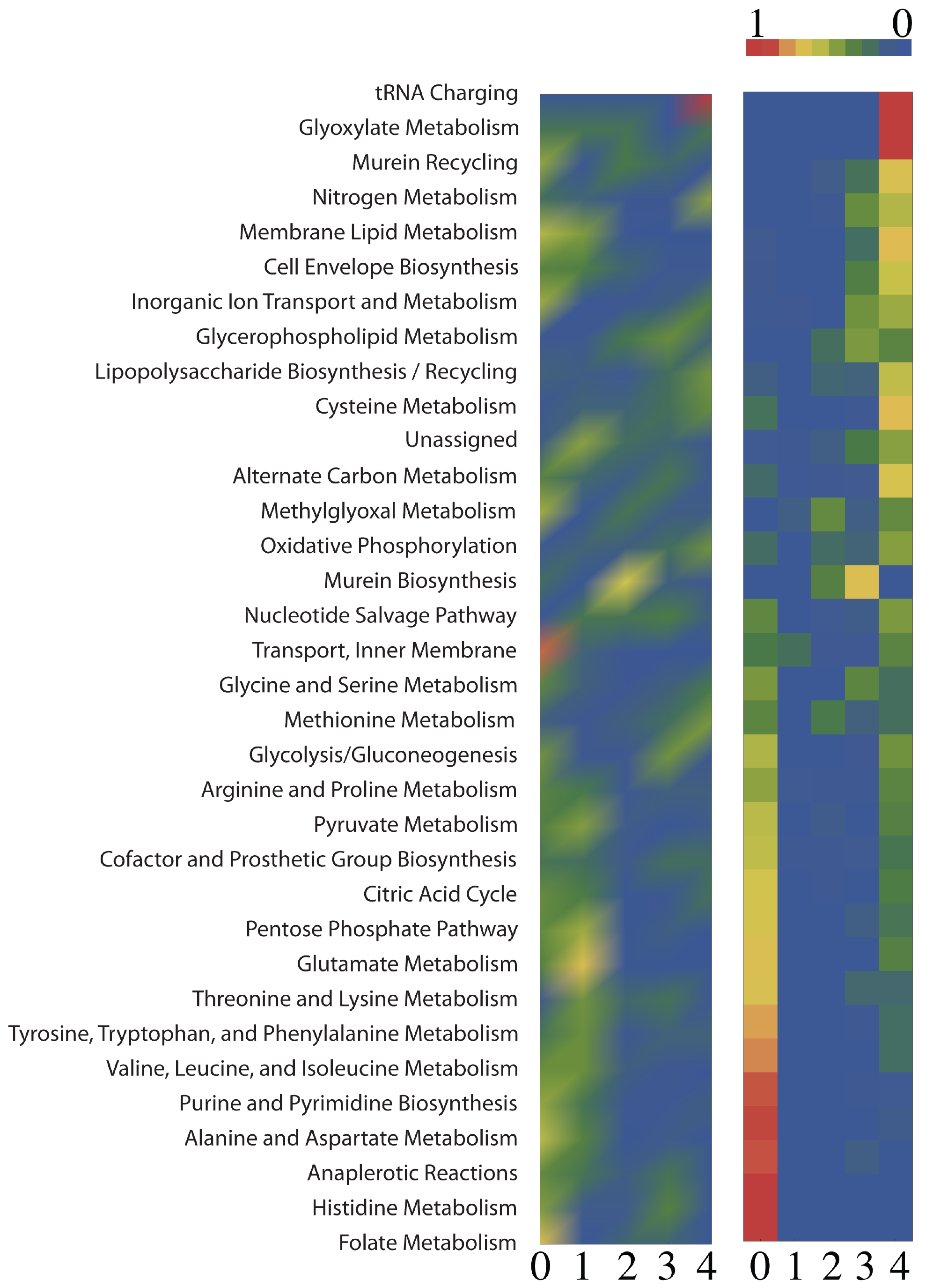} 
\caption{\textbf{Confidence score maps for biochemical pathways in} \textit{E.
coli}\textbf{.} \textbf{Left}. Network-based relative TDB confidence scoring
system. \textbf{Right} Database confidence scoring system (DB). Colors
represent the percentage of reactions within a pathway, as defined
in the database, that have a confidence score with a specific value
(in the $x$ axis). Pathways are ranked from top to bottom according
to the average DB confidence score of reactions in the pathway. We
clarify that the discretization shown for the case of the relative scores $\Sigma$
is just for illustration purposes, and we emphasize that those
are continuous and break the degeneracy inherent to the DB scores.
The used discretization is: value 0 for $\Sigma<1.05$,
value 1 for $1.05\leq \Sigma<1.09$, value 2 for $1.09\leq \Sigma<1.11$,
value 3 for $1.11\leq \Sigma<1.145$, value 4 for $\Sigma>1.145$.}
\label{fig:pathways} 
\end{figure}

Relative scores, $\Sigma$, conform better than absolute scores, $S_{TDB}$,
to the idea of pathways as functional modules since they overexpose
the effect of tree distances, that we expect to be related with the
modular organization of the network. We analyze the $\Sigma$ confidence
score map associated to the different biochemical pathways in \textit{E.
coli} in Fig.~\ref{fig:pathways}. For each pathway, we look at the
distribution of relative scores for the reactions within~\cite{Duarte:2007}. Notice that, even for \textit{E.
coli}, there remain a number of pathways poorly characterized at the experimental level. According to our relative scores, some pathways,
such as \textit{Inorganic Ion Transport and Metabolism} or \textit{Oxidative Phosphorilation}, are associated
to a relatively high average confidence, while others subsystems, such
as \textit{Alanine and Aspartate Metabolism}, present average values
very close to one. At a qualitative level, the overall correlation
between this \textit{in silico} metabolic confidence map and the metabolic
landscape from the DB scoring system is noticeable, although the agreement
is not perfect. In some cases, like for \textit{tRNA Charging}, \textit{Murein
Biosynthesis} or \textit{Glutamate Metabolism}, both maps provide
very good agreement. In contrast, some pathways are at variance, like
the \textit{Transport, Inner Membrane} or \textit{Membrane Lipid Metabolism},
that thus appears as a prominent potential target for further evaluation
and experimental exploration. 

\section{Conclusions} 
The computational network-based TDB confidence scoring system is
able to asses, in probabilistic terms, the reliability of reactions
in metabolic reconstructions solely on the basis of the structure
of the bipartite interactions between metabolites and reactions. It
relies on a link prediction method adjusted to the observations that
exploits the heterogeneity in the number of connections per metabolite
and the hierarchical architecture of the network to estimate connection
probabilites that afterwards are integrated at the level of reactions.
As a result, our TDB scoring system is able to break the degeneracy
of currently employed scores that only use a discrete number of integers
to label different levels of empirical evidence or model-based likelihood,
in addition to providing non-existing information for unranked databases.
TDB scores enable as well further confirmation of steady-state modeling
solutions, uncover very specific reactions that could be functionally
or evolutionary important, and identify prominent experimental targets
for further verification. When compared with a random null model that
justs accounts for heterogeneity in the number of connections per
element, relative scores detect and quantify the tendency of groups
of metabolites to aggregate or disaggregate. This comes out as distance-based
correlations or anticorrelations in the underlying tree metric space,
a question worth exploring in the future in relation to functional
modules.

In a broader context, many biological interactions find a natural
representation in the form of bipartite networks. The ubiquity of
these bipartite structures in cellular networks foretells a wide
range of potential applications of the present methodology, from the
estimation of codon-gene association probabilities to the assessment
of protein complexes.

\begin{acknowledgments}
We thank Mari\'an Bogu\~{n}\'a for helpful discussions. This work was supported by DURSI Project No. 2009SGR1055. F. S. acknowledges financial support by MICINN Project No. FIS2006-03525. M. A. S. was supported by the Ram\'on y Cajal program of the Spanish Ministry of Science and by MICINN Project No. BFU2010-21847-C02-02.
\end{acknowledgments}

\appendix

\section{Fitting the binary tree to the bipartite network} 
We adjust to the observed data a binary tree $T$ with $M+R-1$ internal
nodes at its bifurcation points and $M+R$ leaves corresponding to
the nodes of the metabolic bipartite network, so representing its
$M$ metabolites and its $R$ reactions. Each internal node $t$ has
an associated tree probability $\rho_{t}$ that we transform into
a distance $d_{t}=1-\rho_{t}$. Each pair metabolite-reaction having
as lowest common ancestor $t$ is then separated by a distance $d_{mr}=d_{t}$.
Given a dendrogram, the internal probabilities ${\rho_{t}}$ are only
dependent on the structure of the observed bipartite network and can
be calculated as the fraction of observed connections between leaves
in each branch of the internal node over the total possible. To find
the dendrogram that best fits the real data in terms of likelihood,
we assume that all trees are a priori equally probable and explore
the space of possibilities using a Markov chain Monte Carlo method~\cite{Newman:1999}
combined with a maximum likelihood approach, following the methodology
in~\cite{Clauset:2008}.

In statistical inference, the likelihood $\mathcal{L}$ of a statistical
model for a certain set of observed data is the probability that the
model is a correct explanation, and allows us to estimate its unknown
parameters. For a set of connection probabilities $\rho_{t}$ and
taking into account the underlying tree, the likelihood function becomes
\begin{equation}
\mathcal{L}(D,{\rho_{t}})=\prod_{t\in D}\rho_{t}^{E_{t}}(1-\rho_{t})^{\mathcal{E}_{t}-E_{t}}.\label{lhrgb}\end{equation}
 As for unipartite graphs, the variable $E_{t}$ stands for the number
of actual edges in the observed bigraph, in our case those that connect
metabolites and reactions in the bipartite graph with $t$ as their
lowest common ancestor in $T$. The variable $\mathcal{E}_{t}$ corresponds
to the total possible number of such edges given reactions and metabolites
in the different branches of the common ancestor $t$, discounting
internal combinations. In the unipartite case, $\mathcal{E}_{t}=L_{t}R_{t}$,
being $L_{t}$ and $R_{t}$ the number of leaves in the left and right
subtrees rooted at $t$. In our scheme for bipartite networks, $\mathcal{E}_{t}=L_{tm}R_{tr}+L_{tr}R_{tm}$,
counting possible combinations between metabolite leaves in the left
subtree and reaction leaves in the right subtrees and vice versa. Then, $\rho_t=E_t/\mathcal{E}_{t}$ maximize $\mathcal{L}$.

Starting from a random configuration, we move among all possible sets
of dendrograms by performing random swaps between one of the branches
of a randomly chosen internal node and the alternative branch at its
father level. This exploration is appropriate because it is ergodic
and fulfils detailed balance. The likelihood of the new dendrogram
produced in this way is computed and the dendrogram is accepted or
rejected according to the standard Metropolis-Hastings rule~\cite{Newman:1999}:
the transition is accepted whenever the likelihood does not decrease
and otherwise it is accepted with a probability $\exp(\Delta log\mathcal{L})$
(for computational purposes, it is more convenient to work with the
logarithm of the likelihood function).

After a transient period when $\mathcal{L}$ reaches its equilibrium
value (except typical fluctuations) the system reaches a stationary
state where we sample over $10^{3}$ dendrograms at regular intervals
to produce an average measure of $\rho_{t}$ and so of $\rho_{mr}$
for each possible metabolite-reaction pair. This model, the \textbf{Hierarchical
Random Bipartite Graph} (HRBG), is a generalization for bipartite
networks of the model introduced in~\cite{Clauset:2008}. As explained
in the main text, in our TDB model we correct the tree distances $d_{mr}$
( $d_{mr}=1-\rho_{mr}$) by the heterogeneity in the degrees of metabolites.
We renormalize them according to Eq.~(\ref{prenor}) to produce the
connection probabilities $p_{mr}$ that we use in the construction
of the confidence scores.

\section{Alternative methods} We compare predictions of our TDB model with those produced by other
alternatives like the Configuration Model for Bipartite networks~\cite{Newman:2001b,Newman:2002c,Guillaume:2006}
(CMB), the Hierarchical Random Graph model~\cite{Clauset:2008} generalized
for bipartite networks (HRBG) (values of $\rho_{mr}$ are taken directly
for $p_{mr}$), and a local similarity measure counting Common Neighbors~\cite{Lorrain:1971}
(CN). 

\subsection{The Configuration Model for Bipartite networks} (CMB) assumes
a certain number of reactions $R$, a certain number of metabolites
$M$, and their degree distributions $P(k_{r})$ and $P(k_{m})$,
which should fulfill the requirement $\left<k_{r}\right>R=\left<k_{m}\right>M$,
where $\left<k_{r}\right>$ is the average number of metabolites in
the reactions and $\left<k_{m}\right>$ is the average number of reactions
in which metabolites participate. Metabolites and reactions are partitioned
into two different classes and each element in each class is assigned
an expected degree from the corresponding distribution, which is attached
in the form of stubs. Two stubs, one in each partition, are selected
at random and the link between the metabolite and the reaction is
created avoiding multiple connections.

For the CMB, $p_{mr}=\frac{k_{m}k_{r}}{\left<k_{r}\right>R}$ and Eq.~(\ref{ssc}) can be calculated
analytically. Since the distribution of the bipartite degrees of the reactions is nearly homogeneous, $k_r \approx \left<k_{r}\right>$, it becomes 
\begin{equation}
n_{\nu}\approx R\prod_{m\in\nu}\frac{k_{m}}{R}\prod_{m'\notin\nu}\left(1-\frac{k_{m}'}{R}\right),
\end{equation}
that gives the CMB confidence score for a metabolic reaction when
its set of associated metabolites is $\nu$. 

 \subsection{Common Neighbors} (CN) measure counts the number
of shared neighbors of a given pair. This measure represents a family
of overlap measures quantifying similarity between nodes and a normalized
version was specifically introduced for the study of the hierarchical
modularity of metabolic networks and to delineate the functional modules
based on the network topology~\cite{Ravasz:2002}.In the case of
bipartite metabolic networks, we define common neighbors for a pair
of metabolites as the number of reactions in which they are concurrent,
$o_{mm'}$, and we estimate the probability of connection metabolite-reaction
as $p_{mr}=\sum_{m'\subset r}o_{mm'}/\sum_{\forall m'}o_{mm'}$.

\section{{\it E. coli} bipartite network representation}
In order to build a bipartite network representation of the metabolism of {\it E. coli}, we use the iAF1260 version of the K12 MG1655 strain~\cite{Feist:2007} provided in the BIGG database (http://bigg.ucsd.edu/). It comprises 1039 metabolites and 2381 reactions which include isomerizations, exchange reactions, intracompartment reactions, and different types of transport reactions (some of which involve chemical transformation) between three different compartments: cytosol, periplasm, and a third symbolic one representing the extra-organism.
\begin{figure*}[t]
\begin{center}
\vspace{-1cm}
\includegraphics[width=14cm]{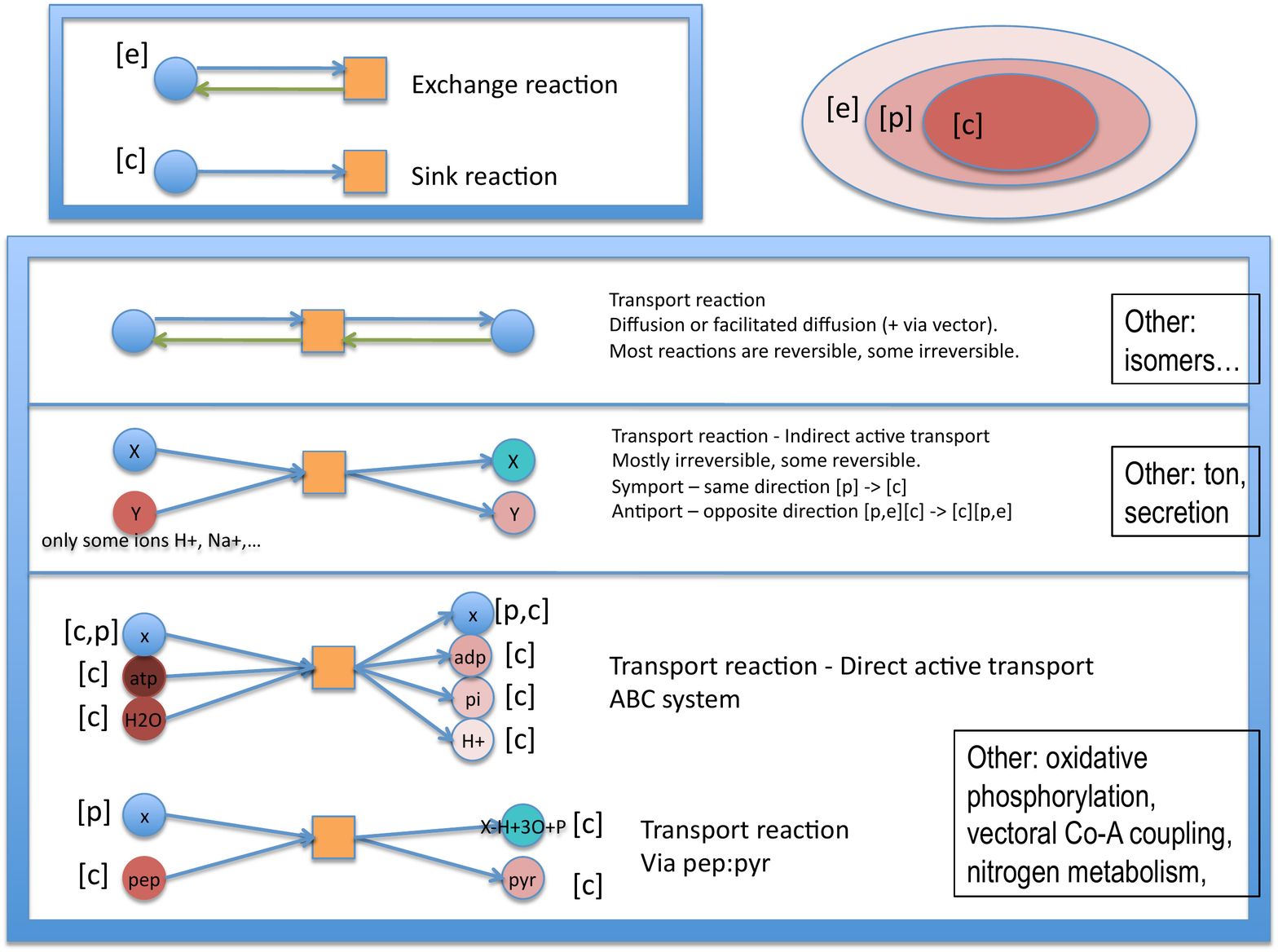}
\end{center}
\caption{Some type of reactions and its topological bipartite representation in E. Coli metabolic network.}
\label{fig:reactions}
\end{figure*}
The most simple representation is in the form of an unweighted undirected network without self-loops or dangling ends (dead end reactions). More refined versions would take into account directionality of the reactions, stoichiometric coefficients, self-loops, etc..

Inside compartments, isomerization reactions (some reversible and some irreversible) transform the structure of one compound without altering its molecular formula. Isomers can have significantly different properties, so in principle it seems reasonable to consider those reactions in a network representation, and the two compounds as separate. However, the topological confidence score of those reactions will depend exclusively on the joint probability of occurrence of the pair of isomers and, most probably, the score would be expected to be low since they only enter together into those reactions. Our option (to be consistent with the treatment of diffusion reactions that lead to the same problematic as we explain below) is to neglect those reactions and to take the isomers as a single entity.

Exchange reactions represent the exchange of metabolites between the cell and the environment. These are reversible and involve a single metabolite. In a bipartite representation, exchange reactions would be dangling ends with a single incoming connection. Sinks needed to allow metabolites to leave the system (irreversible reactions in the cytoplasm) can be treated in the same way. Since they would count in the total amount of reactions but never two metabolites would enter them simultaneously, it seems reasonable to neglect them in the context of this work. Furthermore, exchanges or sinks have no associated reconstruction confidence score in the database.

Regarding transport reactions, there are several options. In {\it E. coli}, three different compartments are differentiated: cytosol, periplasm, and extra-organism. Every compound in a different compartment is considered as an individual specie and transport steps are formally considered as reactions transferring the compound belonging to one compartment into the same compound belonging to the other compartment (the respective concentrations can be different, and the compartments usually have different volume). Transport reactions between compartments are of different kind. Basic general mechanisms are:
\begin{itemize}
  \item Via diffusion (membrane permeable to water molecules and a few other small, uncharged, molecules like
oxygen and carbon dioxide that diffuse freely in and out of the cell) or via facilitated diffusion (transmembrane proteins create a water-filled pore through which ions and some small hydrophilic molecules can pass by diffusion. The channels can be opened or closed according to the needs of the cell). Molecules and ions move spontaneously down their concentration gradient. No chemical transformation is involved. In the database, most are reversible processes while some are irreversible.
  \item Via active transport. Active transport is defined as mass transport from a region of lower to a region of  a higher electrochemical potential.  It is transport that requires energy.  Active transport across biological membranes occurs via enzymes. Transmembrane proteins, called transporters, use the energy of ATP (hydrolyzation of ATP, in general $ATP + H2O -> ADP + Pi$, $\Delta G^0_{ATP}=-30,3KJ/mol$ or $-7.3 kcal/mol$, enough to pump 2 sodium ions) to force ions or small molecules through the membrane against their concentration gradient. Active transport can be direct or indirect:
  \begin{itemize}
    \item Direct active transport. Some transporters bind ATP directly and use the energy of its hydrolysis to drive active transport (ATP powered pumps). In the database, these are named as ABC (ATP-Binding Cassette) transmembrane proteins, which expose a ligand-binding domain usually restricted to a single type of molecule at one surface and an ATP-binding domain at the other surface; the ATP bound to its domain provides the energy to pump the ligand across the membrane. In the database, hydrolyzation of ATP correspods to the chemical transformation $atp[c] + h2o[c] -> adp[c] + h[c] + pi[c]$ (cytoplasm, periplasm).
    \item Indirect active transport. A discrete class of proteins import or export ions and small molecules, such as glucose and amino acids, against a concentration gradient. These proteins use the energy stored in the electrochemical gradient of a directly-pumped ion to power the uphill movement of another substance, a process called cotransport. Direct active transport of the ion establishes a concentration gradient. When this is relieved by facilitated diffusion, the energy released can be harnessed to the pumping of some other ion or molecule. Symport pumps: the driving cotransported ion (H+,Na+) and the transported molecule pass through the membrane pump in the same direction. The driving ions flow down their concentration gradient while the coupled molecules are pumped up theirs; later the ion is pumped back out of the cell by a direct active transport process.  Antiport pumps: the driving ion (again, usually sodium, proton/phosphate/succinate/nitrite/) diffuses through the pump in one direction providing the energy for the active transport of some other molecule or ion in the opposite direction. In any case, no direct chemical change in the database.
  \end{itemize}
\end{itemize}
In the database, other transport reactions also appear. Some do not involve chemical transformation like for indirect active transport: via ton system, or secretion (without known transport mechanism). A second class involve chemical transformation of metabolites along transport: pep-pyr system, oxidative phosphorylation, vectorial Co-A coupling, nitrogen metabolism, etc..

In relation to the bipartite network representation, one needs to differentiate the different transport mechanism: diffusion, transport without chemical transformation, and transport involving chemical transformation of carriers (see Fig.~5).

Diffusion reactions do not transform any compound and involve the same metabolite both as input and output (self-loops in a metabolite one-mode projection), and so require some specific treatment. One possibility is to consider the metabolites in both parts of the reaction as different entities, the effective pair. Since this is the only pair entering such reactions, the topological score is expected to be low as for isomerization reactions. However, since the metabolite is the same with the same properties, a better option in this case is to neglect these reactions and do not differentiate the metabolite in different compartments.

Other transport reactions without chemical transformation involve more than one metabolite, usually two that change compartment. In topological terms, this is a situation analogous to diffusion and the same treatment of ignoring the reaction and taking the metabolite in different compartments as the same can be applied.

Transport reactions that need to transform metabolites to transfer a compound have different metabolites entering and leaving the reaction except the compound that is transported across compartments. To be consistent with the previous treatments, that metabolite in different compartments would be treated as a single entity. 

Finally, chemical transformation reactions that happen in more than one compartment (or that need some component of a different compartment from where it is happening) can be treated as chemical transformation reactions inside compartments.

\begin{figure}[t]
\begin{center}
\includegraphics[width=8.7cm]{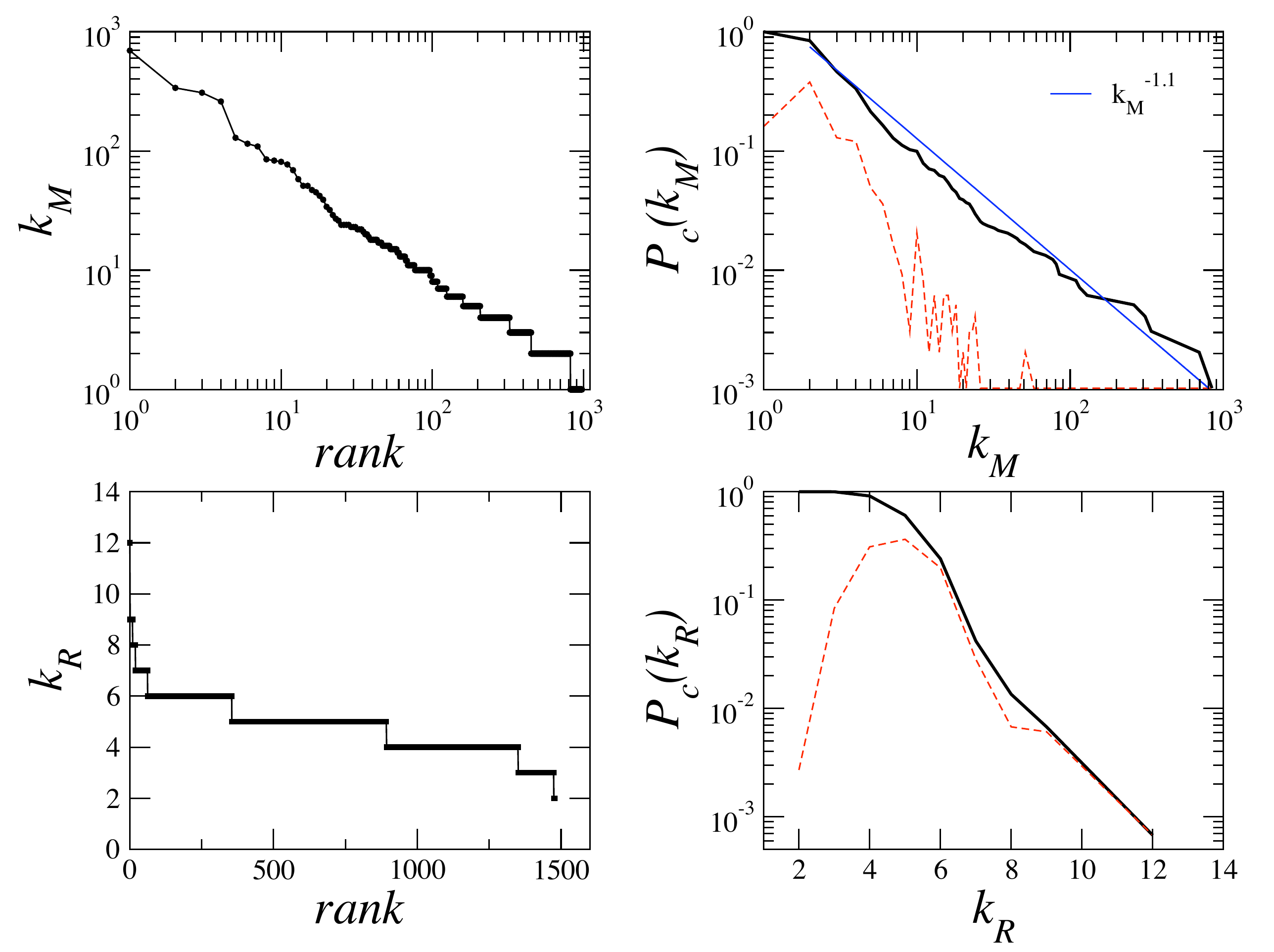}
\end{center}
\caption{{\bf Rank distribution of degrees of metabolites and reactions and cumulative bipartite degree distributions.} The degree of a metabolite is taken as the number of reactions it participates in and the degree of a reaction is the number of different metabolites it involves.}
\label{fig:degrees}
\end{figure}
\begin{figure}[t]
\begin{center}
\includegraphics[width=8.7cm]{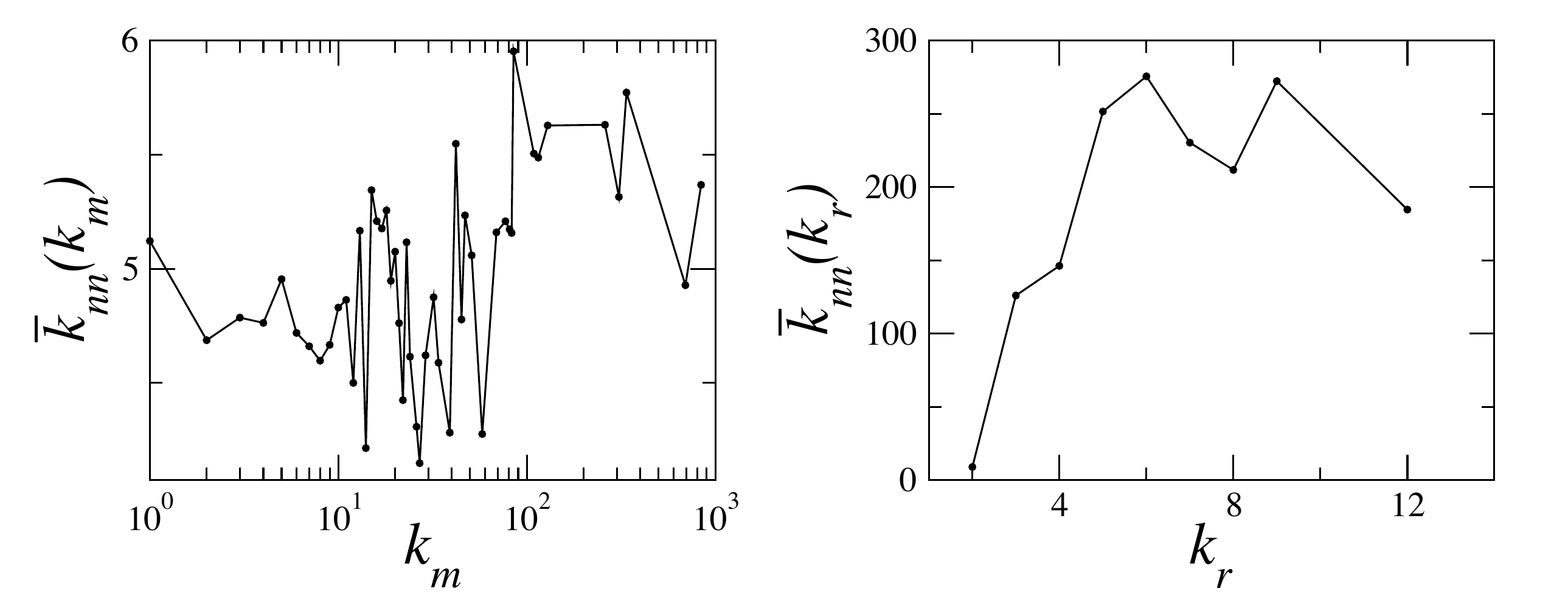}
\end{center}
\caption{{\bf Average nearest neighbors bipartite degree of metabolites and reactions.} }
\label{fig:degrees}
\end{figure}
In summary (see Table~1), we take metabolites as the same entity independently of the compartment and we obviate isomerizations, exchange, and diffusion reactions. This leaves a total of $1479$ reactions out of $2381$ that involve $976$ metabolites out of $1039$ (isomers have been identified with the same id). Apart from the identification of isomers, 5 metabolites have been removed ($mn2$ (id 30), $ca2$ (id 60), $na1$ (id 383), $ag$ (id 651), and $cl$ (id 654)) because they do not enter in any transformation reaction. They seem to be necessary to transfer compounds (typically $h$ but also others) across compartments.
\begin{table*}[h]
\caption{Classification of reactions in the database according to compartment implications, function and topological structure. The total number of reactions $R$ in the database is 2381.}
\begin{center}
\begin{tabular}{@{} lcp{8cm}rr}
\hline \hline\\
\multicolumn{5}{c}{Classification of reactions in E. Coli metabolic network}\\
Type&ID&Description& Number of reactions& \hspace{0.5cm}Included in graph\\[0.3cm] \hline \hline \\[0.1cm]
\multirow{5}{*}{internal}&1& cytosol & 1100&Yes\\
&2& isomerization in cytoplasm & 59&No\\ &3& periplasm & 190&Yes\\ &4& isomerization in periplasm& 2&No \\
&5& extra-organism & 8&Yes\\  \\[0.1cm] \hline  \\[0.1cm]
\multirow{2}{*}{exchange}&6& sinks in cytosol & 5&No\\&7& exchange in extra-organism& 299&No\\ \\[0.1cm] \hline \\[0.1cm]
\multirow{7}{*}{transport}&8& diffusion, facilitated diffusion, via vector, channel, flipping& 346&No\\&9& ABC system, direct active transport (+1 detoxification)&124 &Yes\\ &10& symport, +1 similar  reaction ID 11690
& 124&No\\&11&antiport, +6 similar reactions Ids 11407, 11409, 11411, 11413, 11415, 11835 & 50&No\\
&12&ton system & 11&No\\&13&secretion (transport mechanism not known) & 6&No\\
&14&reactions with transformation involving different compartments (pep:pyr, oxidative phosphorylation, vectoral Co-A coupling, nitrogen metabolism, etc.) & 57&Yes\\ \\[0.1cm]
\hline \hline
\end{tabular}
\end{center}
\label{table_types}
\end{table*}

In Fig. 6, we show the cumulative bipartite degree distribution both for metabolites and reactions. The degree of a metabolite is taken as the number of reactions it participates in and the degree of a reaction is the number of different metabolites it involves.  The number of metabolites entering into a reaction $k_r$ follows a homogeneous distribution with mean $<k_r>=4.82$ and mode $5$. In contrast, the number $k_m$ of reactions in which a metabolite participates displays a scale free degree distribution $P(k_m)\sim k_m^{-2.1}$ with an average degree $<k_m>=7.30$. Currency metabolites are the most connected substrates, some with more than a hundred and up to $841$ connections ($h^+$, $h2o$, $atp$, $pi$, $adp$, $ppi$, $nad$, $nadh$).

The average nearest neighbors bipartite degree of metabolites and reactions is displayed in Fig. 7. The rising of the average levels for higher values of the degrees means that the reactions involving more metabolites also involve more carrier metabolites (hubs).
compartments and also enter in exchange reactions.




\end{document}